\documentclass[times, 12pt,epsf]{article}
\usepackage{times}
\usepackage{psfig}

\begin{document}

\title{\bf A Game Theoretic Economics Framework to understanding the Information Security outsourcing Market}

\author{
Wen Ding \hspace{1in} William Yurcik\\
National Center for Supercomputing Applications (NCSA)\\
University of Illinois at Urbana-Champaign\\
\{wending,byurcik\}@ncsa.uiuc.edu}

\maketitle

\thispagestyle{empty}

\begin{abstract}

On information security outsourcing market, an important reason
that firms do not want to let outside firms(usually called
MSSPs-Managed Security Service Providers) to take care of their
security need is that they worry about service quality MSSPs
provide because they cannot monitor effort of the MSSPs. Since
MSSPs action is unobservable to buyers, MSSPs can lower cost by
working less hard than required in the contract and get higher
profit. In the asymmetric information literature, this possible
secret shirking behavior is termed as moral hazard problem. This
paper considers a game theoretic economic framework to show that
under information asymmetry, an optimal contract can be designed
so that MSSPs will stick to their promised effort level. We also
show that the optimal contract should be performance-based, i.e.,
payment to MSSP should base on performance of MSSP's security
service period by period. For comparison, we also showed that if
the moral hazard problem does not exist, the optimal contract does
not depend on MSSP's performance. A contract that specifies
constant payment to MSSP will be optimal. Besides these, we show
that for no matter under perfect information scenario or imperfect
information scenario, the higher the transaction cost is, the
lower payment to MSSPs will be.

\end{abstract}


\vspace{0.1in}

\section{Introduction} \label{sec:intro}

Security outsourcing market where firms contract with outside
information security vendors to meet their organizational demands
has been growing at a double digit rate for the past $3$ years,
and experts predict that this growth rate will continue through
$2008$\cite{DeSo04}. Compared with the booming of the business,
theory of security outsourcing is less developed. In view of this
both buyers and MSSPs need to strategically understand the nature
of this market.

Information security outsourcing is different from traditional
outsourcing because information security is different from durable
goods and other services outsourced such as payroll and
accounting. As more and more firms automate processes, servers and
the networks work like the brains and vessels of a firm. If any
core system go down, the cost may be large due to lost data and
lost revenue. What makes it worse is that security breaches are
irreversible. While defects in manufacturing can be returned or
wrong paychecks can be reissued, monetary loss due to down time is
gone forever, and lost customer confidence may be hard to gain
back. Therefore, while most industries put cost saving as the
primary reason they outsource business processes other than
security, firms that outsource information security state service
quality is their primary motivation. This is supported by a survey
by Jeffrey Kaplan published in Business Communication Review
($2003$)\cite{Kapl03}. It is reported that 40.6\% of the firms
outsource network operations based on concerns for service
quality.

Information asymmetry is another reason that firms have concerns
outsourcing their security. Since buyers cannot observe and
monitor MSSPs' action, MSSPs, as profit maximizing companies, have
an incentive to lower their effort level to reduce cost.

The model we present is a model where buyers and MSSPs engage in a
repeated game with infinite horizon where MSSPs' effort level in
not observable to buyers. We show that under this information
asymmetry, moral hazard problem will occur. Performance based
contracts are recommended to avoid such moral hazard problem.

For comparison, we also provide results under perfect information,
where buyers can have all information they need and shirking is
not an option for MSSPs. Under the scenario of perfect
information, the optimal solution(in terms how the contact is
written) is a price-only contract. This solution is called first
best because no deadweight loss is incurred under perfect
information assumption.

Besides the optimal contract form, we are particularly interested
in the effect of transaction cost on market equilibrium price.
Transaction cost includes all cost spent on searching for, arguing
and executing contracts with MSSPs\cite{Coas37}. We argue in
section(\ref{sec:theocost}) transaction cost  can be very high in
outsourcing non traditional services such as security because
standard rules and procedures have not been established yet. We
show that when transaction cost increases, price of security
outsourcing will be lowered.

There is a large body of literature on IT outsourcing, including
information security outsourcing as a sub-category. Ang and Straub
(1998)\cite{AnSt98} did an empirical study on the U.S. banking
industry and showed IT outsourcing is strongly influenced by the
production cost advantage offered by IT service vendors.
Transaction cost also influences outsourcing decisions with a much
smaller effect. Though their result is based on data of US banking
system, this result is probably true in a lot of areas outside the
banking system. Based on their result, we will assume decrease in
production cost out-weight increase in transaction cost throughout
this paper. Lacity and Willcocks'(1998)\cite{LaWi98} use US and UK
organizations survey data and provide empirical evidence that the
following practices are recommended to achieve cost saving
expected: selective outsourcing, senior executives and IT manager
make decisions together, invite both internal and external bids,
short-term contract, detailed fee-for-service contract. This paper
will provide theoretical support for the last practice. Mieghem
(1999)\cite{VanM99} builds a game theoretic model on production
outsourcing where investment decision has to be made before market
demand is revealed. After market demand is revealed, the firm's
production is limited to its investment level, and will use
outside production(outsource) to meet excess demand. His paper
studies three kinds of contracts 1), price-only contract,
2),incomplete contract and 3), state-dependent contract. He shows
that only state-dependent contract is optimal in the sense that it
eliminates all decentralizing cost\footnote{centralized economy
system assumes there is a social planner who make decision by
pooling all available resources from different firms.
Decentralized economy system is one where firms make their own
decision using individual resources. It can be shown that outcome
of centralized economy {weakly} dominates outcome of decentralized
economy. Difference between the two is decentralization cost.} His
paper is related to security outsourcing because in security
outsourcing, an implicity assumption of centralized economy is
that all participants will work diligently. Therefore, with moral
hazard problem, decentralization cost is caused by the possibility
that MSSPs may shirk. This paper will investigate why
state-contingent contract is preferred to non state-contingent
contract from a information economics point of view. We argue that
state-contingent contract is the optimal contract form when there
is moral hazard problem.

The rest of this paper is organized as follows: In Section 2 and
3, we contrast information security outsourcing with other types
of outsourcing.  Next we set up an outsourcing model with perfect
and imperfect information to discuss what optimal contract look
like and what is the effect of transaction cost on prices in
Section 4. In Section 5, related work on this topic is summarized.
We end with a summary and conclusions in Section 6.

\thispagestyle{empty}

\section{Outsourcing Theory} \label{sec:theory}

Outsourcing is defined as `all the subcontracting relationships
between firms and the hiring of workers in non-traditional jobs'
(Heshmati 2003)\cite{Hesh03}. Business Process Outsourcing (BPO),
which includes outsourcing of human resources, finance and
accounting, procurement, shared services, billing, customer care
and so on, is estimated to grow at a 9.5\% compound annual rate
through 2007 reaching \$173 billion by Gartner\cite{SDSH03}. IT
Outsourcing (ITO) is expected to grow at a compound rate of 7.2\%
through 2008 reaching \$253.1 billion in 2008\cite{CYGS04}.
Furthermore, Information security outsourcing is predicted to grow
from \$4.1 billion in 2001 to \$9.0 billion in 2006, a compound
growth rate of double digits\cite{DeSo04}.

Behind this booming of outsourcing, the basic force is `cost
efficiency'. As markets become more competitive, outsourcing is an
essential way firms may reduce costs. By using information
security outsourcing, firm only need to pay a fraction of their
in-housing cost for outsourced security. Outsourcing can reduce
cost either because suppliers has lower input costs and/or larger
scale of production as in the case of offshore manufacturing
outsourcing; or because the suppliers have expertise or more
advanced technology as in payroll and IT outsourcing. However, at
the same time of reducing production cost, buyers incur
transaction costs\cite{Coas37} searching for, signing, and
executing contracts with suppliers. In the case of total
outsourcing, when firms keep no in-house production, firms also
lose sunk costs\footnote{Firm's investment specific to the
outsourced process}, which can be machines and plants that can
only be used to produce the outsourced product or can be money
spent on training technicians.

If cost reduction is the only concern for firms, firms will
outsource when reduction in production cost exceeds increase in
transaction cost. In standardized outsourcing procedures such as
payroll and manufacture goods, transaction cost has been reduced
as Coase\cite{Coas37} predicted `This(transaction) cost may be
reduced but it will not be eliminated by emergence of
specialist$\ldots$'. It is argued that transaction cost is some
percentage of the contract value since the larger the project, the
greater effort firms will spent on searching for a proper MSSP and
the more coordination is needed between firm and MSSP after
signing the contract.

The second outsourcing incentive is firms will be able to
concentrate on their core competence by outsourcing
support/routine functions. For example, although a lot computer
companies are based in the U.S., most keyboards are produced in
Asia. By outsourcing labor intensive processes to areas that are
abundant in labor, firms achieve cost reduction and become more
focused on core competence.

Yet another key reason for outsourcing is to obtain higher
quality. Outside companies accumulate more experience by
specializing in certain processes. They can afford larger
investment on R\&D to get updated technology and skills and better
trained expertise. A large client base also contributes to the
quality of goods and services of outside producers and service
providers. They gain experience and knowledge by serving varied
clients. Consulting, for example, the service providers have
professional knowledge that a non-consulting firm can never afford
to build by itself.

Argument against production outsourcing concerns unemployment
issue as in off-shore outsourcing: while argument against security
outsourcing focus on transaction cost control and service quality
monitoring. We will analyze these two concerns on information
security outsourcing in detail in the following section.

\thispagestyle{empty}

\section{Security Outsourcing: What is Special?}
\label{sec:secout}

In spite of all the advantages outsourcing may bring, some people
think security should not be outsourced, or firms should be really
careful when doing so.

\subsection{Quality Measurement Difficulty}

Security management is an art rather than science where we know
how to achieve a best solution; here we do not even know what the
best solutions are, nor do MSSPs. A security system can be a very
complicated project. People may think that they are safe with
firewalls and IDSs. Even so, firms have to decide which firewalls
and IDSs to buy, how to allocate limited budget on combination of
these devices to reach maximum level of security and how to manage
these devices and tune them so that they secure your system enough
and do not give too many alerts on harmless behaviors. The bright
side is MSSPs are gaining experience on these issues quickly by
their devotion and specialization in this area.

However, people argue that it is hard to evaluate products and
services of MSSPs both ex ante and ex post. As security
outsourcing market becoming prominent over the last few years; a
large number of MSSPs emerged from diversified backgrounds. The
largest ones include firms formed solely to solve internet
security problems such as Counterpane, firms from research and
computer production such as IBM, anti virus companies such as
Symantec, firms from internet providers such as AT\&T and so on.
This diversification in background reflects on their diversified
product and services making it really hard for the firms to
compare and choose from them. (See appendix I for major MSSPs and
their products.)

Also, evaluating MSSPs' products by performance of their products
is tricky because the outcome is highly random and can even be
misleading. A better secured system may be down because of
intensive attacks; systems that ignore patching notices from time
to time may go well for a long time. On the other hand, it is not
true that the more money spent on security, the fewer bleaches a
system will have. Sophisticated hackers are more attracted to
systems that are hard to break into.

However, a 'better' secured system should be less vulnerable in
statistical sense in the long run. This paper will use
\underline{\emph{expected}} performance to evaluate a security
system. We assume buyers have access to historical data of MSSP's
service performance, and can generate a distribution of benefit
from using security outsourcing.

\subsection{Effective Cost Reduction?}\label{sec:theocost}

Based on a survey on IT managers, directors and other decision
makers from both firms that outsourced security and those who did
not, cost reduction remains their focus\cite{Kapl03}.

There is evidence that security outsourcing will reduce production
cost. Device management for example, which tunes and monitors
firewalls, IDSs and runs vulnerability testing, a security
personnel cost\$8,000 to \$16,000 per month. And to get 24*7
support, this figure may need to be more than tripled. For the
same functions, MSSPs charge between \$600 and \$4,000. For
network monitoring, Counterpane, one of the most successful MSSPs,
claims that it only charges a fraction of the money for net
management a firm need to spend to do the security in house: `From
an annualized basis, its going to cost you \$1 million to \$1.2
million just to look at the sam information we monitor, and our
average contract ranges from \$40,000 to \$150,000 a year ---
between 4\% and 10\% of what it would cost to do yourself
$\ldots$'\cite{Mill04}.

However, although security vendors' may provide huge reduction in
production cost, transaction cost may be quite high. Since
standard measure for security services has not been established
and each MSSP uses their featured(different) technology, most of
the time it is very hard to do comparison across different MSSPs.
This quality measurement difficulty may increase transaction cost
potentially\cite{PoZe98}.

Also, writing up the contract and decide who is responsible for
what kind of losses due to security breaches can be painful. Firms
would feel more comfortable if security vendors can take
responsibility if losses occur. But it is not always the security
vendor's fault because no matter how well security devices are
designed and tuned, there is always probability that the system is
broken into. More tricky things can be if security vendors take
responsibility for the losses, firms may not play due diligence as
they should. Therefore, although this paper is devoted to
discussion of MSSPs' moral hazard behavior, the optimal contract
needs to guard against firms' moral hazard behavior as well, which
may increase transaction cost significantly. Therefore although we
will assume that transaction cost is lower than reduction in
production cost, effect of transaction cost needs to be further
explored.

\thispagestyle{empty}

\section{The Model} \label{sec:model}

Based on above observation of how security outsourcing is special,
We set up the model in the following way.

There are two sides on the security outsourcing market: potential
security service buyers (``buyers" for short), and security
vendors(MSSPs). Vendors and buyers all seek to maximize their
individual profit.

Basic assumptions are:
\begin{itemize}
   \item A1: Vendors are more cost efficient than firms; transaction cost
   is lower than production cost advantage.
   \item A2: Services provided by different security vendors are imperfect
substitutes\footnote{imperfect substitutes are goods that are not
identical but have similar functions, e.g. lap-top and desk-tops.
}.
    \item A3: Buyers do not have moral hazard problem.
\end{itemize}

In the following three subsections, we show that:
\begin{enumerate}
\item With imperfect information, we have moral hazard problem on
MSSP side. Optimal contract depends non-trivially on MSSPs
performance.

\item With perfect information, optimal contract is a price-only
contract.

\item With either perfect information or imperfect information,
price is decreasing on transaction cost.
\end{enumerate}

\subsection{Optimal contract with imperfect information\\ --- Performance based contract}\label{sec:imp}

Due to imperfect information, actions of the players are not
directly observable. Both MSSPs and security buyers can disobey
their promises secretly. In this paper, we focus on how to avoid
moral hazard behavior of MSSPs, and assume buyers will always
follow the contract as it is. The optimal contract will be such
that following the contract is the best choice for both players.
We temporarily assume transaction cost is zero in this section.

Our analysis is based on principal-agent problem with infinite
horizon following Spear and Srivastava()\cite{SpSr87}, where
agent's action is not observable to principal. principal is
assumed to be risk neutral\footnote{A risk neutral player only
cares about average payoff.}and agent risk averse\footnote{A risk
averse player gets lower utility if variance of his payoff
increase}. Here, MSSP is agent to principal buyer. We are allowed
to assume security buyer is risk neutral because security buyers
have access to insurance market and can buy insurance to mitigate
risks that MSSPs cannot eliminate. However, the risk neutral
assumption is not essential to the result. We can discuss risk
averse buyers but it only make the mathematics more complicated
without accomplishing anything. So we just keep the simple
assumption that buyers are risk neutral.

Denote buyer's period t benefit(before payment to MSSP) from
security outsourcing as $y_t$. Because of the random nature of
cyber attacks, $y_t$ is a random variable. Denote MSSP's effort
level in period t as $a_t$, $a_t\in[\underline{a},\overline{a}]$.
Then distribution of security service performance $y_t$ is
conditional on MSSP's effort $a_t$. Denote the distribution as
$f(y,a_t)$. $P_t$ denotes buyer's compensation(price) to MSSP in
period t. History up to period t is denoted as:
$h_t=\{y_t,y_{t-1},\ldots,y_0\}$.

A price contract is composed of MSSP's effort level and price
buyer pays to MSSP: $\{a_t(h_{t-1}), P_t(h_t)\}$. Notice that
MSSP's period t effort level $a_t$ depends only on history up to
period t-1, since MSSP has to choose his effort level at beginning
of period t before period t benefit $y_t$ is realized. Payment to
MSSP in period t however depends on the whole performance history.

Let $u(P_t)-\phi(a_t)$ be net payoff to MSSP under contract
$\{a_t(h_{t-1}), P_t(h_t)\}$, where $u(P_t)$ is MSSP's utility
from payment $P_t$ and $\phi(a_t)$ measures cost of working at
effort level $a_t$. We assume $u'>0$, $u''<0$\footnote{$u''<0$
comes from risk averse assumption.} and $\phi'>0$. History $h_t$
evolve recursively by the following probability rule:
\begin{eqnarray}
\pi(h_t|h_{t-1})=f(y_t,a_t(h_{t-1}))\pi(h_{t-1})
\end{eqnarray}

Assume buyers and MSSP discount future payoff at same rate $\rho,
\rho\in[0,1]$, then buyer and MSSP's period t expected payoff are
$\int(y_t-P_t)f(y_t|a_t)dy_t$ and $u(P_t)-\phi(a_t)$:

Discount all future payoff to period 0, we have buyer and MSSP's
period 0 discounted payoff as:
\begin{eqnarray}
B_t(P_t, a_t)&=&\sum_{j=0}^{\infty}\sum_{h^{t+j}}\rho^j
[\int(y_t-P_t)f(y_t,a_t)dy_t]\pi(h_{t+j},a_{t+j}|h_t)\\
M_t(P_t,a_t)&=&\sum_{j=0}^{\infty}\sum_{h^{t+j}}\rho^j
[u(P_t)-\phi(a_t)]\pi(h_{t+j},a_{t+j}|h_t)
\end{eqnarray}

Therefore, the maximization problem for security buyer is to
choose a sequence of contracts $\{P_t(y), a_t\}_{t=0}^\infty$ to
maximize discounted expected utility subject to the constraint
that MSSP cannot benefit from deviating from the contract:
\begin{eqnarray}
\max_{\{P_t(y)\}_{t=0}^\infty, \{a_t\}_{t=0}^\infty}&\quad& B_t(P_t(y), a_t)\nonumber\\
\mbox{st}&\quad& M_t(P_t(y),a_t)\geq M_t(P_t(y),\tilde{a}_t) \quad
\forall \tilde{a}_t\in[\underline{a},\overline{a}]\label{equi:IC}
\end{eqnarray}

where, constraint in above maximization problem is called the
incentive compatibility(IC) constraint. It show that the effort
level $a_t$ is optimal for MSSP compared to any other possible
effort level $\tilde{a}_t$.

Since the above problem has infinitely unknown variables, it is
impossible to solve it directly. Instead, we rewrite it in the
recursive form.

In the recursive form, principal maximize current period's payoff
assuming he will behave optimally from next period on. Let $v$
denote payoff buyer promised to MSSP this period and $w(y)$ denote
the promised payoff to MSSP next period. $K(v)$ be maximized
payoff to buyer when MSSP gets v as promised expected payoff.
Hence, $K(w(y))$ is buyer's best possible payoff next period. Then
the maximization problem in recursive form is:

\begin{eqnarray}
K(v)&=\quad&\max_{P(y),w(y),a}\quad \int[y-P(y)+\rho K(w(y))]f(y,a)dy\nonumber\\
\mbox{st} &\quad& \int[u(P(y))+\rho w(y)]f(y,a)dy-\phi(a)\geq
v\quad \quad \quad \quad \mbox{(PK)}\nonumber\\
& & a\in \arg\max \int[u(P(y))+\rho w(y)]f(y,a)dy-\phi(a) \quad
\mbox{(IC)}\label{sys:rec}
\end{eqnarray}

The optimal contract should contain $\{P(y),w(y),a\}$. (PK) is
short for ``promise keeping''. It requires that if buyer promised
MSSP payoff v, the contract should guarantee expected payoff to
MSSP is at least v(equal to v in equilibrium). (IC) constraint is
same as in (\ref{equi:IC}).

The (IC) constraint implies the solution $a$ should satisfy both
the following first order condition and second order condition:

\begin{eqnarray}
(FOC)& &\quad \quad\int[u(P(y))+\rho w(y)]f_a(y,a)dy-\phi'(a)\\
(SOC)& &\quad \quad\int[u(P(y))+\rho
w(y)]f_{aa}(y,a)dy-\phi''(a)\leq 0 \quad\quad \forall w(y)
\end{eqnarray}

Assumption:
\begin{itemize}
    \item Convexity of distribution function condition(COFC):
    \begin{eqnarray}
    F_{aa}\geq 0
    \end{eqnarray}
    where $F(x,a)=\int_{-\infty}^x f(y,a)dy$
\end{itemize}

Rogerson(1985)\cite{Roge85} shows that when COFC is satisfied,
(SOC) is guaranteed. We can use (FOC) to substitute (IC)
constraint and get rid of the (SOC).

Let $\lambda$ be Lagrangian multiplier on (PK) constraint and
$\mu$ be the multiplier on (IC)-(FOC) constraint. We have the
Lagrangian equation:
\begin{eqnarray}
L&=&\int[y-P(y)+\rho K(w(y))]f(y,a)dy\nonumber\\
& &+\lambda(\int[u(P(y))+\rho w(y)]f(y,a)dy-\phi(a)-v)\nonumber\\
& &+\mu(\int[u(P(y))+\rho w(y)]f_a(y,a)dy-\phi'(a))
\end{eqnarray}

Take first order conditions w.r.t $P(y),w(y)$ and $a$, we get the
following first order conditions and the envelope condition:
\begin{eqnarray}
\{P(y)\}& &\quad\quad -1+\lambda u'(P(y))+\mu u'(P(y))\frac{f_a(y,a)}{f(y,a)}=0\label{equi:foc1}\\
\{w(y)\}& &\quad\quad \rho
K'(w(y))+\rho\lambda+\mu\rho\frac{f_a(y,a)}{f(y,a)}=0\label{equi:foc2}\\
\{a\}& &\quad\quad \int[y-P(y)+\rho P(w(y)]f_a(y,a)dy \nonumber\\
& &\quad \quad \quad\quad+\mu[\int[u(P(x))+\rho
w(y)]f_{aa}(y,a)dy-\phi''(a)]=0\label{equi:foc3}\\
\{ENV\} & & \quad\quad K'(v)=-\lambda\label{equi:ENV}
\end{eqnarray}

First order conditions (\ref{equi:foc1}) and (\ref{equi:foc2})
implies:

\begin{eqnarray}
\frac{1}{u'(P(y))}=-K'(w(y))=\lambda+\mu\frac{f_a(y,a)}{f(y,a)}\label{equi:equ}
\end{eqnarray}

Definition: MLRP(monotone likelihood ratio property)
\begin{itemize}
    \item Likelihood ratio $\frac{f_a(y,a)}{f(y,a)}$ is monotone
    in $y$
    or $\frac{d}{dy}[\frac{f_a(y,a)}{f(y,a)}]\geq 0$. This also
    implies: $\forall a>\tilde{a}, y>\tilde{y},
    \frac{f(y,a)}{f(\tilde{y},a)}\geq\frac{f(y,\tilde{a})}{f(\tilde{y},\tilde{a})}$.
    \end{itemize}

Intuitively, this means at a higher effort level $a$, it is more
probable to get a higher benefit $y$ than at a lower effort level
$\tilde{a}$.

Rogerson(1085)\cite{Roge85} shows that when the density function
$f(y,a)$ has monotone likelihood ratio property, $\mu$ the
multiplier on (IC) constraint is positive.

When MLRP holds, $\mu>0$, equation (\ref{equi:equ}) implies the
following results:
\begin{description}
    \item[Result 1]
    $y\uparrow\Rightarrow\frac{1}{u'(P(y))}\uparrow\Rightarrow
    P(y)\uparrow$.\\
    Reason: $u''(P(y))\leq 0$\\
    This result suggests contacts should be performance-based, i.e. payment to MSSP
    should be higher when benefit from security outsourcing increases and
    vice versa. And this supports empirical result of Lacity and Willcock(1998)\cite{LaWi98}.
    \item[Result 2]$y\uparrow\Rightarrow K'(w(y))\downarrow\Rightarrow
    w(y)\uparrow$\\ Reason: $K(w(y))$ is best possible payoff of buyer next period
    when MSSP's expected payoff is w(y). Since MSSP's payoff comes
    from compensation $P(y)$ from buyer, the higher MSSP's payoff
    $w(y)$ is, the lower buyer's payoff$K(w(y))$ will be.\\
    This result suggest buyer should reward MSSP with higher expected payoff
    for next period if buyer gets high benefit this period.
    \item[Result 3]$v\uparrow\Rightarrow\lambda\uparrow\Rightarrow
    P(y),w(y)\uparrow$\\Reason:
    $v\uparrow\Rightarrow\lambda\uparrow$ from the envelope
    condition (ENV). $\lambda\uparrow\Rightarrow
    P(y),w(y)\uparrow$ follows from equation (\ref{equi:equ}).\\
    This result shows that if buyer promise MSSP a higher current
    expected payoff, buyer should increase both current period
    compensation and next period promised expected payoff.
\end{description}

To sum up, from Result 1 - 3, we suggest that optimal contract
under moral hazard should depend on performance in a non-trivial
way. And effect of performance is persistently on future
compensations. The effect is carried over by promised value $v$
and $w(y)$ as shown in Result 2 and 3.

\subsection{Optimal contract with perfect information \\--- price only contract}\label{subsec:per}

With perfect information, buyer can monitor MSSP's behavior very
well. Then MSSP is not able to shirk and moral hazard problem does
not exist. In this scenario, Maximization problem of
buyer(\ref{sys:rec}) reduces to:
\begin{eqnarray}
K(v)&=\quad&\max_{P(y),w(y),a}\quad \int[y-P(y)+\rho K(w(y))]f(y,a)dy\nonumber\\
\mbox{st} &\quad& \int[u(P(y))+\rho w(y)]f(y,a)dy-\phi(a)\geq
v\quad \quad \quad \quad \mbox{(PK)}
\end{eqnarray}

Corresponding first order conditions are:
\begin{eqnarray}
\{P(y)\}& &\quad\quad -1+\lambda u'(P(y))=0\label{equi:foc4}\\
\{w(y)\}& &\quad\quad \rho K'(w(y))+\rho\lambda=0\label{equi:foc5}\\
\{a\}& &\quad\quad \int[y-P(y)+\rho P(w(y)]f_a(y,a)dy=0\label{equi:foc6}\\
\{ENV\} & & \quad\quad K'(v)=-\lambda\label{equi:ENV}
\end{eqnarray}

Equation {\ref{equi:foc4}} and (\ref{equi:foc5}) imply:
\begin{eqnarray}
\frac{1}{u'(P(y))}=-K'(w(y))=\lambda\label{equi:equ1}
\end{eqnarray}

This suggests that without moral hazard problem, optimal
compensation and next period promised value does not depend on
this period's outcome $y$. Constant compensation and promised
value would be optimal.

\subsection{Effect of transaction cost}

\subsubsection{Effect from game between buyer and MSSP}

In this section, we will study how transaction cost affects
equilibrium market price. No matter whether buyer has perfect
information about MSSP's effort level or not, existence of
transaction cost reduces buyers compensation to MSSP.

As in section(\ref{sec:imp}), we use $P(y)$ to denote buyer's
compensation to MSSP. Since buyers will also need to pay
transaction cost on top of service price, the actual out of pocket
price buyers of MSSP face is $(1+\alpha)P(y)$, where $\alpha P(y)$
is the transaction cost\footnote{transaction cost is modelled as a
percentage of contract value because as the project gets larger,
buyer and vendor need to spend more time and money on the
negotiation and coordination part \cite{Coll04}. A Survey done by
Barthelemy(2001)\cite{Bart01} shows that transaction cost is up to
6\% for contracts lower than \$10million value}.

With transaction cost, we modify the maximization problem of buyer
as:
\begin{eqnarray}
K(v)&=\quad&\max_{P(y),w(y),a}\quad \int[y-(1+\alpha)P(y)+\rho K(w(y))]f(y,a)dy\nonumber\\
\mbox{st} &\quad& \int[u(P(y))+\rho w(y)]f(y,a)dy-\phi(a)\geq
v\quad \quad \quad \quad \mbox{(PK)}\nonumber\\
& & a\in \arg\max \int[u(P(y))+\rho w(y)]f(y,a)dy-\phi(a) \quad
\mbox{(IC)}\end{eqnarray}

Corresponding first order conditions are:
\begin{eqnarray}
\{P(y)\}& &\quad\quad -(1+\alpha)+\lambda u'(P(y))+\mu u'(P(y))\frac{f_a(y,a)}{f(y,a)}=0\label{equi:foc7}\\
\{w(y)\}& &\quad\quad \rho
K'(w(y))+\rho\lambda+\mu\rho\frac{f_a(y,a)}{f(y,a)}=0\label{equi:foc8}\\
\{a\}& &\quad\quad \int[y-P(y)+\rho P(w(y)]f_a(y,a)dy \nonumber\\
& &\quad \quad \quad\quad+\mu[\int[u(P(x))+\rho
w(y)]f_{aa}(y,a)dy-\phi''(a)]=0\label{equi:foc9}\\
\{ENV\} & & \quad\quad K'(v)=-\lambda\label{equi:ENV1}
\end{eqnarray}

From first order conditions (\ref{equi:foc7}) we have 
\begin{eqnarray}
\frac{1+\alpha}{u'(P(y))}=\lambda+\mu\frac{f_a(y,a)}{f(y,a)}\label{equi:equ2}
\end{eqnarray}

Similarly, under perfect information, we have:
\begin{eqnarray}
\frac{1+\alpha}{u'(P(y))}=\lambda\label{equi:equ3}
\end{eqnarray}

Compare with equation (\ref{equi:equ}) and equation
(\ref{equi:equ1}), it can be implied that all other things same,
compensation $P(y)$ is smaller with transaction cost.

\subsubsection{Effect from game among MSSPs}
Another effect of transaction cost on market price comes from
competition among MSSPs. This effect also suggests when
transaction cost increase, nominal market price will decrease.
\begin{itemize}
   \item A3: Vendors engage in a price competition against each other.
\end{itemize}

We will derive the Nash Equilibrium\footnote{A strategy vector x
with payoff vector $\pi$ is called a Nash Equilibrium if
$\pi_i(x_i, x_{-i})\geq \pi_i(\tilde{x_i}, x_{-i}),\forall
\tilde{x_i}\in X_i, \forall i$. $X_i$ is set of all possible
actions player $i$ can take. This condition means that Nash
Equilibrium is such that no player can benefit from unilateral
deviations.}\cite{Nash50} price under the assumption A1-A3. For
this section, to see effect of MSSPs' competitions, we ignore
effect of buyers, and assume perfect information(as shown in
section(\ref{subsec:per}), optimal contract specifies a
non-performance-dependent price, $P(y)$ is replaced with $P$). We
will show that MSSPs will lower price to bear part of the
transaction cost due to competition with other MSSPs. Division of
the transaction cost between buyers and vendors depends on demand
elasticity for security products.

A price competition is where every MSSP uses price as a strategic
variable, and is free to choose a price that maximizes their
profit given price of other vendors. Explicitly, profit
maximization problem for vendor i is:

$$
\max_{P^i}\{P^i\cdot N^i((1+\alpha)P)- C^i(N^i((1+\alpha)P)\}
$$

$P$ denotes the price vector $\{P^i, i=1,\ldots , V\}=\{P^i,
P^{-i}\}$, where $P^i$ is market price MSSP$i$ charges. $P^{-i}$
is the price vector of prices of all other MSSPs except MSSP$i$
charges. $N^i$ is demand for MSSP$i$'s service, which depends on
market prices. It also depends on service quality MSSPs provide
implicitly. $C^i$ is MSSP$i$'s total cost of servicing $N^i$
customers. Then the above maximization problem shows how MSSP$i$
maximize its net profit(revenue minus cost) by choosing $P^i$ when
other vendors charge price $P^{-i}$.

$C^i$ includes both fixed cost($FC$) which does not change with
number of customers and variable cost($VC$) which does.
Explicitly,
\begin{equation}
C^i(N^i(\cdot))=FC + VC(N^i(\cdot)),
\end{equation}

$C(\cdot)$ increases with number of customers.

Optimal price MSSP$i$ should charge solves the following first
order condition of the maximization problem w.r.t $P^i$:
\begin{equation}
N^i(\cdot)+P^i\frac{\partial N^i(\cdot)}{\partial
P^i}(1+\alpha)=C'(N^i(\cdot))\frac{\partial N^i(\cdot)}{\partial
P^i}(1+\alpha) \label{equi:opti1}
\end{equation}

Divide both sides of equation (\ref{equi:opti1}) with
$\frac{\partial N^i(\cdot)}{\partial P^i}(1+\alpha)$ and rearrange
terms, we get:
\begin{equation}
P^i(1-\frac{1}{\eta^i(1+\alpha)})=C'(N^i(\cdot))\quad \mbox
i=1,\ldots, V \label{equi:opti2}
\end{equation}
where $\eta^i=-(\partial{N^i(\cdot)}/N^i)/(\partial{P^i}/P^i)$,
which represents percentage change in demand due to percentage
change in price, the price elasticity of vendor i's demand. It
measures how sensitive market demand changes with price. Because
$\partial d(\cdot)/\partial(P)<0$(demand and price move in
opposite directions), a negative sign is added so that $\eta>0$.

solving $P^i$ from optimizing condition (\ref{equi:opti2}), $P^i$
is a function of $P^{-i}$, $\alpha$ and $\eta$:

\begin{equation}
P^i=r(P^{-i}, \alpha, \eta)\label{equi:opti3}
\end{equation}

Equation(\ref{equi:opti3}) can be viewed as response function of
MSSP $i$ on prices of other security MSSPs $P^{-i}$. Therefore,
for all MSSPs on the market, $i=1,\ldots, V$, we can form a
equation system:

\begin{eqnarray}
P^1=r(P^{-1}, \alpha, \eta)\nonumber,
\\ P^2=r(P^{-2}, \alpha, \eta)\nonumber,
\\ \ldots \nonumber
\\ P^V=r(P^{-V}, \alpha, \eta)\label{equi:opti4}
\end{eqnarray}

The Nash Equilibrium of this price competition is a price vector
(\emph{strategies}) that solves the above equation system and a
corresponding vector of profit(\emph{payoffs}). Under regularity
conditions, this equilibrium price vector exists and is
unique\cite{Nash50}.

To give an idea how this Nash Equilibrium price look like, we
present a graphic solution for the simplified case when $V=2$.
Then optimization conditions (\ref{equi:opti4}) reduce to the
following:

\begin{eqnarray}
P^1=r(P^2, \alpha, \eta)\nonumber
\\ P^2=r(P^1, \alpha, \eta)
\label{equi:two0}
\end{eqnarray}

To make things easier, we make two more assumptions:
\begin{itemize}
   \item A4. Marginal cost $C_i'(\cdot)$ is constant, i.e. it
costs MSSP $i$ same amount of money to serve one additional buyer.
   \item A5. $\frac{\partial \eta^i}{\partial (P^i/P^{-i})}>0$,
meaning, as MSSP $i$'s service becomes more expensive relative to
services of other MSSPs, demand for MSSP $i$'s service become more
elastic. In other word, a same percentage increase in $P^i$ will
induce greater percentage reduction in $N^i$ for higher
$P^i/P^{-i}$ then lower.
\end{itemize}

\begin {figure} [!h]
\begin{center}
\begin{minipage}[b] {2.3in}
\begin{center}
\centerline{\psfig{figure=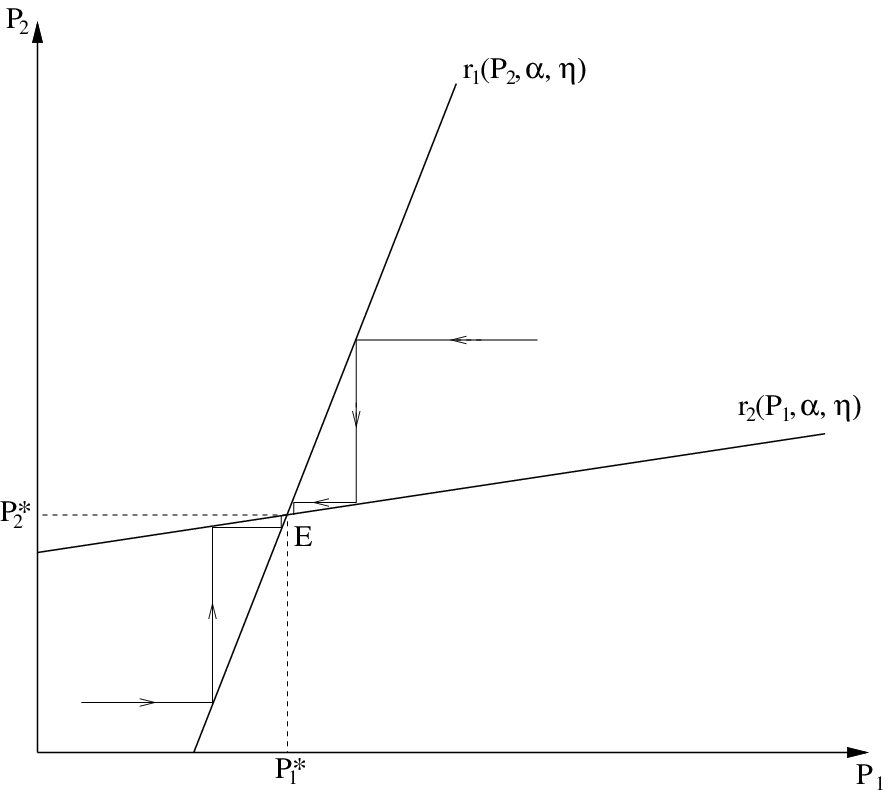,width=2.3in}} {\small (a) Nash
Equilibrium prices when $\alpha=0$}
\end{center}
\end{minipage}
\begin{minipage}[b] {2.3in}
\begin{center}
\centerline{\psfig{figure=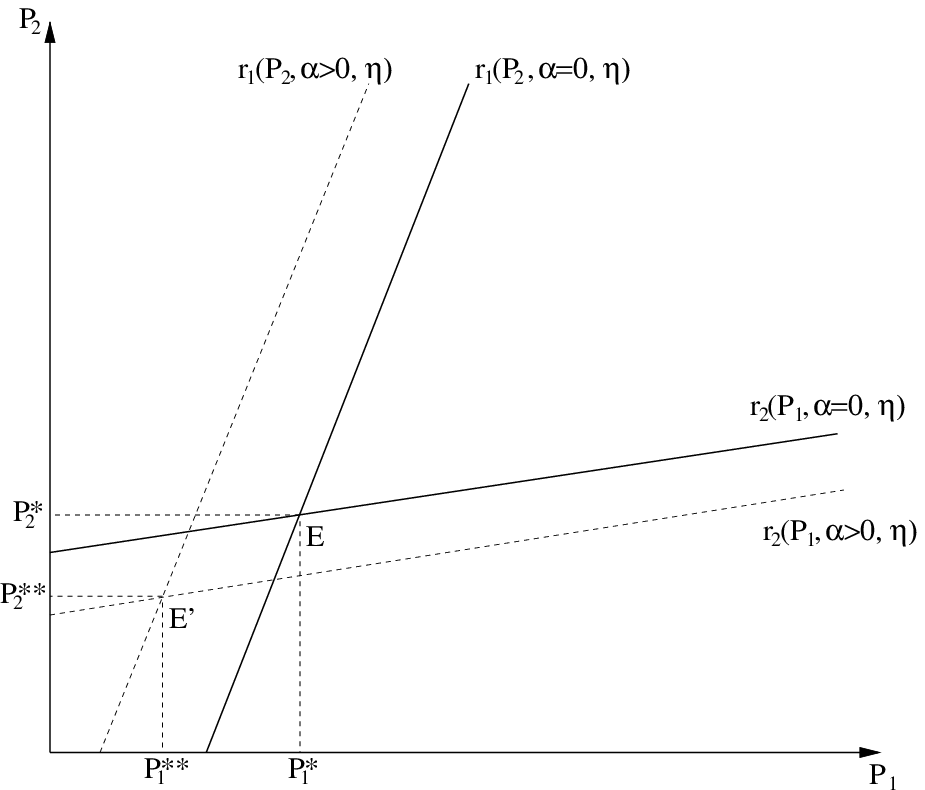,width=2.3in}} {\small (b) Nash
Equilibrium prices when $\alpha>0$}
\end{center}
\end{minipage}
\end{center}
\caption{Effect of transaction cost on equilibrium price}
\label{fig:equil}
\end {figure}

Two response curves $P^i=r(P^{-i},\alpha, \eta), i=1,2$ are
plotted in figure 1 where the horizontal axe represent MSSP 1's
price and the vertical axe represent MSSP 2's price. Under A4 and
A5, Feenstra\cite{Feen04} showed that both reaction curves have
positive slopes. Then slope of MSSP 1's response curve is larger
than slope of that of MSSP 2's as shown in Fig.\ref{fig:equil}(a).

Because response curve is the locus of MSSP's best responses given
the other MSSP's action, the intersection point E is the
equilibrium point where both MSSPs are are choosing optimally and
simultaneously. By definition, they are the Nash Equilibrium
prices. Observe that this Nash Equilibrium is a stable equilibrium
in the sense that no matter what price the MSSPs start off with,
they will eventually arrive at point E, as shown by the arrows in
Fig.\ref{fig:equil}(a).

Denote price vendor $i$ would charge by $P^i_0$ when there is no
transaction cost($\alpha=0$), from equation system
(\ref{equi:two0}),
\begin{eqnarray}
P^i_0=r(P^{-i}, \alpha=0, \eta)
\quad, i=1,2
\end{eqnarray}

Totally differentiate optimization condition (\ref{equi:opti2}),
\begin{eqnarray}
dP^i(1- \frac{1}{\eta^i(1+\alpha)})+P^i \frac{d
\eta^i}{\eta^{i2}(1+\alpha)} +P^i\frac{d
\alpha}{\eta^i(1+\alpha)^2}=C''(N(\cdot)) \label{equi:total}
\end{eqnarray}
By A4
\begin{eqnarray}
C''(N(\cdot))=0
\end{eqnarray}
Equation (\ref{equi:total}) implies:

\begin{eqnarray}
dP^i(1-\frac{1}{\eta^i(1+\alpha)}+\frac{\frac{d\eta^i/\eta^i}
{dP^i/P^i}}{\eta^i(1+\alpha)})=-P^i\frac{d\alpha}{\eta^i(1+\alpha)}
\label{equi:total1}
\end{eqnarray}

Assume:

\begin{itemize}
    \item A6. $\frac{d\eta^i/\eta^i}{dP^i/P^i}>1-\eta^i(1+\alpha)$
    \label{assu:ineq}
\end{itemize}
Under assumption (\ref{assu:ineq}),
\begin{eqnarray}
1-\frac{1}{\eta^i(1+\alpha)}+\frac{\frac{d\eta^i/\eta^i}
{dP^i/P^i}}{\eta^i(1+\alpha)}>0
\end{eqnarray}

Equation(\ref{equi:total1}) implies
\begin{eqnarray}
d\alpha>0\quad\Rightarrow\quad dP^i<0
\end{eqnarray}

This shows that when transaction cost increases, MSSPs reduce
their prices correspondingly.

Graphically, the reaction curve $P^1=r(P^2, \alpha, \eta)$ shifts
to the left and $P^2=r(P^1, \alpha, \eta)$ shifts down. therefore,
compare with the reaction curves when there is no transaction
cost. As shown in Figure-\ref{fig:equil}(b), reaction curves with
transaction cost intersect at lower price level for both MSSPs.
Remember that the intersection of reaction curves is the Nash
Equilibrium of the game.

As shown above, under assumptions 1-6, existence of transaction
cost reduces prices charged by MSSPs. The extend of reduction
depends on how sensitive market demand is to prices.

\section{Related Work} \label{sec:relwk}

\subsection{Empirical Work}

Empirical works on this issue were mostly done with surveys. Ang
and Straub (1998) performed a well designed survey on banks of
different sizes with items measuring degree of IT outsourcing,
production cost advantage, transaction cost, financial slack
(archive data also used here) outsourcing degree and firm size.
And they found that production cost advantage is the main driving
force of IT outsourcing, transaction cost dampens outsourcing
intention, but has a much smaller effect. They also reported
evidence that degree of IT outsourcing decreases with firm size.
They argued that this is because large firms are more likely to
generate economies of scale in their IT department, therefore are
more likely to produce IT services in-house. Lacity and Willcocks
(1998) measures success or failure of a IT outsourcing based on
seven factors, and found that outsourcing scope, length of
contract term, contract type are among the most important factors
that decides how successful an IT outsourcing is. Poppo and Zenger
(1998) includes technological uncertainty, measurement difficult
and quality satisfaction in their model, and showed that when it
is harder to measure performances, firm become less satisfied with
costs. Ang and Cummings (1997) found empirical evidence that in
hyper-competitive environments, not only firms act strategically,
but security vendors also.

\subsection{Analytical Work}

Analytical papers on the other hand have a strong game theoretic
flavor. Mieghem (1999) built a multivariate, multidimensional
competitive model, and investigated effect of subcontracting
complexity on coordination.

Ang and Cummings argued that organizations respond strategically
under hyper-competitive environments. Whang employed a game
theoretical approach to explain asymmetric information and
incentive compatible issue in software development.

\section{Conclusion}

Security outsourcing market benefits both vendors and buyers if it
works properly. In the first place, security outsourcing offers
cost reduction for buyers. We showed that for security
outsourcing, optimal form of contract should be performance-based.
Also, we showed that with transaction cost, price paid to MSSPs
are lower than otherwise. MSSPs take part of the transaction cost
to stimulate demand.

\thispagestyle{empty}

\bibliographystyle{plain}
\bibliography{../ref}

\thispagestyle{empty}

\end{document}